\documentclass[11pt]{article}
\usepackage{jheppub,amsmath, amsthm, amssymb,slashed,url}
\usepackage{graphicx}
\usepackage{epstopdf}
\usepackage{tikz}
\usepackage{xcolor}
\usepackage{tabu}
\usepackage{tabularx}

\usepackage{etoolbox}
    \makeatletter
    \patchcmd{\maketitle}{\@fpheader}{}{}{}
    \makeatother

\usepackage{boldline}
\usepackage{bm}
\usepackage{amssymb,amsfonts}
    
\def\sss{\scriptscriptstyle}

\def\oneht{\textstyle{1\over 2} }

\def\onefourth{\textstyle{1\over 4} }

\def\OMIT#1{{}}

\def\si{^1 \hskip -0.03in S _0}
\def\siii{^3 \hskip -0.025in S _1}

\newcommand{\beq}{\begin{equation}}
\newcommand{\eeq}{\end{equation}}
\newcommand{\beqa}{\begin{eqnarray}}
\newcommand{\eeqa}{\end{eqnarray}}

\newcommand{\nn}{\nonumber}

\begin{document}

\dedicated{NT@UW-21-14, IQuS@UW-21-015}
\title{Causality and dimensionality in geometric scattering}
\vskip 0.5cm
\author{\bf Silas R.~Beane}
\author{\bf and Roland C.~Farrell}
\affiliation{InQubator for Quantum Simulation (IQuS), Department of Physics,\\
  University of Washington, Seattle, WA 98195.}

\vphantom{} \vskip 1.4cm

\abstract{The scattering matrix which describes low-energy,
  non-relativistic scattering of spin-1/2 fermions interacting via
  finite-range potentials can be obtained from a geometric action
  principle in which space and time do not appear
  explicitly~\protect\cite{Beane:2020wjl}. In the case of zero-range forces, 
  causality leads to constraints on scattering trajectories 
  in the geometric picture.  The effect of spatial dimensionality 
  is also investigated by considering scattering in two and three dimensions. 
  In the geometric formulation it is found that dimensionality is encoded in the phase 
  of the harmonic potential that appears in the geometric action.}

\maketitle


\section{Introduction}

\noindent The scattering matrix ($S$-matrix) encodes observable
consequences of the quantum mechanical interaction of two particles.
While the $S$-matrix is typically obtained by solving an effective 
quantum field theory (EFT) ---usually in
perturbation theory--- the end-product $S$-matrix is strictly a
function of on-shell kinematical variables. Therefore, the effects 
of space and time, which are naturally embedded in any EFT calculation 
of the $S$-matrix, are entirely encoded in the kinematical variables. 
Motivated by the non-local correlations between the scattered 
particles due to quantum entanglement, a question one might ask is: 
can the $S$-matrix be obtained in a
spacetime-independent formulation without direct reference to an EFT\footnote{For recent efforts
  in this direction from the perspective of perturbative gauge
  theories, see
  Refs.~\cite{Arkani-Hamed:2020gyp,Arkani-Hamed:2019vag}.}?

Recent work~\cite{Beane:2020wjl} by the authors has answered the above
question in the affirmative by providing a geometric construction of
the non-relativistic scattering of spin-1/2 fermions\footnote{This
  simple system, with the fermions identified as nucleons, has the
  advantage of being the low-energy EFT of the Standard Model at
  distances greater than the Compton wavelength of the pion, and has
  widespread and important applications in nuclear physics. See
  Ref.~\cite{Beane:2000fx,Hammer:2019poc} for a review.}.  In the geometric theory,
the $S$-matrix is a trajectory in the space of all phase shifts
allowed by unitarity, and is parameterized by a kinematical
energy-momentum variable that is determined up to Galilean
transformations.  A critical observation in Ref.~\cite{Beane:2020wjl}
is that the $S$-matrix allows a momentum inversion symmetry which is
not manifest in the EFT of contact operators\footnote{The inversion
  symmetry does manifest itself in the renormalization group (RG)
  evolution of the EFT couplings~\cite{Beane:2021A}.}. This symmetry maps 
  low- and high-energy scattering processes into each other and so is a UV/IR symmetry. 
  This UV/IR symmetry leaves classes of
observables invariant and is conformal in the sense that 
various combinations of phase shifts are left unchanged. In the geometric
picture this corresponds to reflection symmetries of the $S$-matrix
trajectories. The UV/IR symmetry allows the construction of exact
solutions in the geometric theory, including specification of the
harmonic forces which deform the geodesics in the geometric
space. These exact solutions provide a forum for the exploration of
spacetime constraints on the $S$-matrix in the geometric formulation
of scattering.

The $S$-matrix evolves a state vector from the boundary of space in
the infinite past, to the boundary of space in the infinite future,
and must do so in a manner consistent with causality and with
awareness of the number of spatial dimensions in which it is acting.
Constraints due to causality on non-relativistic scattering have
implications for the analytic structure of the
$S$-matrix~\cite{PhysRev.74.131,PhysRev.83.249,PhysRev.91.1267} and,
for systems arising from finite-range potentials, for the range of
allowed values of the scattering
parameters~\cite{Wigner:1955zz,Phillips:1996ae,Hammer:2009zh,Hammer:2010fw}.
These bounds, known as Wigner bounds, provide powerful constraints on
the exact $S$-matrix solutions implied by conformal invariance. In the
geometric theory it is found that these bounds manifest themselves as
constraints on the tangent vectors of scattering trajectories. 
 In addition, as quantum mechanics depends strongly on spatial
dimensionality, the differences between scattering in two and 
three dimensions are explored in the geometric formulation. 
The resulting $S$-matrix in two spatial dimensions again
has a solution implied by conformal invariance. Despite the strikingly
different physics that it gives rise to, the form of the
two-dimensional harmonic potential differs from its three-dimensional
counterpart only by a change of coupling strength and phase.

This paper is organized as follows.  Section~\ref{sec:smatth} sets up
the $S$-matrix framework, focusing on the properties of the most
general $S$-matrix consistent with finite-range forces. The $S$-matrix
is shown to allow conformal symmetries that are not manifest in the
EFT action, and which provide powerful geometric constraints. In
Section~\ref{sec:geom}, the $S$-matrix of contact forces is shown to
be the solution of a dynamical system which evolves the two-particle
state in a two-dimensional space defined by the two phase shifts and
bounded by unitarity. The conformal symmetries allow an exact
determination of the forces that determine the $S$-matrix in this
space. These first two sections expand on material first
presented in Ref.~\cite{Beane:2020wjl}.  Section~\ref{sec:stcon} is
new material which explores the manner in which spacetime features of
scattering manifest themselves in the geometric theory of
scattering. Constraints due to causality are considered, and the
dependence on spatial dimensionality is found by varying between three
and two dimensions.  
Finally, Section~\ref{sec:conc} summarizes and concludes.

\section{$S$-matrix theory}
\label{sec:smatth}

\subsection{$S$-matrix theory of contact forces}

\noindent It is a simple matter to write down the $S$-matrix without
reference to any underlying field theory by directly imposing general
physical principles and symmetries. Consider two species of
equal-mass, spin-$1/2$ fermions, which we label as neutrons and
protons (i.e. nucleons), that interact at low energy via forces that are
strictly of finite range. The spins of the two-body system can be
either aligned or anti-aligned. Therefore, near threshold the
$S$-matrix is dominated by the s-wave and can be written as~\cite{Beane:2018oxh}
\begin{eqnarray}
  \hat {\bf S}(p)
  & = &
{1\over 2}\left(  e^{i 2 \delta_1(p)} + e^{i 2 \delta_0(p)} \right)
 \hat   {\bf 1}
\ +\
{1\over 2}\left( e^{i 2 \delta_1(p)} - e^{i 2 \delta_0(p)} \right)
\hat{{\cal P}}_{12}
  \label{eq:Sdef}
\end{eqnarray}
where the SWAP operator is
\begin{eqnarray}
\hat{{\cal P}}_{12} = \oneht \left({ \hat   {\bf 1}}+  \hat  {\bm \sigma} \cdot   \hat  {\bm \sigma} \right)
  \label{eq:swap}
\end{eqnarray}
and, in the direct-product space of the nucleon spins,
\begin{eqnarray}
 \hat {\bf 1} \equiv \hat {\cal I}_2\otimes  \hat {\cal I}_2 \ \ \ , \ \
  \hat {\bm \sigma} \cdot \hat {\bm \sigma} \equiv \sum\limits_{\alpha=1}^3 \
\hat{{\sigma}}^\alpha \otimes \hat{{ \sigma}}^\alpha \ ,
  \label{eq:Hil}
\end{eqnarray}
where ${\cal I}_2$ is the $2\times 2$ unit matrix, and the $\hat{{
    \sigma}}^\alpha$ are the Pauli matrices. The $\delta_s$ are s-wave
phase shifts with $s=0$ corresponding to the spin-singlet ($\si$)
channel and $s=1$ corresponding to the
spin-triplet ($\siii$) channel.

The SWAP operator takes an initial unentangled product state, say
$\lvert p \uparrow \rangle \, \lvert n \downarrow\rangle$, and
scatters it into the unentangled product state $\lvert p \downarrow
\rangle \, \lvert n \uparrow\rangle$. In general, the $S$-matrix is an
entangling operator as the basis which diagonalizes the interaction is
different from the single-particle basis which describes the initial product
state.  Therefore, it is 
imperative to treat the $S$-matrix as the fundamental object of study,
rather than the EFT action or the scattering amplitude, when
addressing issues related to quantum entanglement.

The entangling character of the $S$-matrix is captured by its
entanglement power (EP)~\cite{Beane:2018oxh,Beane:2020wjl,Low:2021ufv}
\begin{eqnarray}
{\mathcal E}({\hat {\bf S}})
& = &
{1\over 6}\ \sin^2\left(2(\delta_1-\delta_0)\right)
\ ,
  \label{eq:epSnf2}
\end{eqnarray}
which is a state-independent measure of the entanglement generated by
the $S$-matrix acting on an initial product state. Note that this
object manifestly couples the two spin states in a manner that is
quite distinct from the Lorentz-invariant, spin decoupled interactions
that are encoded in the EFT action. Indeed, when it vanishes there is
an enhanced $SU(4)$ spin-flavor symmetry (Wigner's supermultiplet
symmetry~\cite{Wigner:1936dx,Wigner:1937zz,Wigner:1939zz}) which
explicitly relates the singlet and triplet scattering channels.

The two angular degrees of freedom, the phase shifts $\delta_{0,1}$,
are characterized by the scattering lengths, $a_{0,1}$ and
effective ranges $r_{0,1}$, near threshold via the effective range
expansion
\begin{eqnarray}
  p\cot\delta_s(p) & =& -\frac{1}{a_s} \ +\ \oneht r_s p^2  \ +\ {\mathcal O}(p^4)
  \label{eq:EFT1}
\end{eqnarray}
with ${\vec p}$ (with $p=|{\vec p}\,|$) chosen to be the center-of-mass (c.o.m.) momentum.
The omitted ${\mathcal O}(p^4)$ corrections are known as shape parameters.
Near threshold, the $S$-matrix can be written as
\begin{eqnarray}
  \hat {\bf S} & = & \frac{1}{2}\left( S_1+ S_0 \right) \;\hat {\bf 1} \;+\; \frac{1}{2}\left( S_1- S_0 \right)\;\hat{{\cal P}}_{12} \ ,
   \label{eq:Sdefres}
\end{eqnarray}
where the $S$-matrix elements are 
\begin{eqnarray}
S_s &=& e^{2i\delta_s(p)} \ =\ \frac{1- i a_s(p) p}{1+ i a_s(p) p} \ ,
   \label{eq:Selementscatt}
\end{eqnarray}
and a momentum-dependent scattering length is defined as
\begin{eqnarray}
a_s(p) \ \equiv\ \frac{a_s}{1-\oneht a_s r_s p^2 + {\mathcal O}(p^4)} \ .
   \label{eq:mdepscatt}
\end{eqnarray}
In terms of phase shifts
\begin{eqnarray}
\phi & \equiv& 2\delta_0\ =\ -2\tan^{-1}\!\left( a_0(p) p \right) \ \ \ , \ \ \ \theta\ \equiv\ 2\delta_1\ =\ -2\tan^{-1}\!\left( a_1(p) p \right) \ .
 \label{eq:confRmodSOL}
\end{eqnarray}
Here the phase shifts have been expressed in terms of the angular variables $\phi\in [0,2\pi]$ and $\theta\in [0,2\pi]$.

\begin{figure}[!ht]
\centering
\includegraphics[width = 0.45\textwidth]{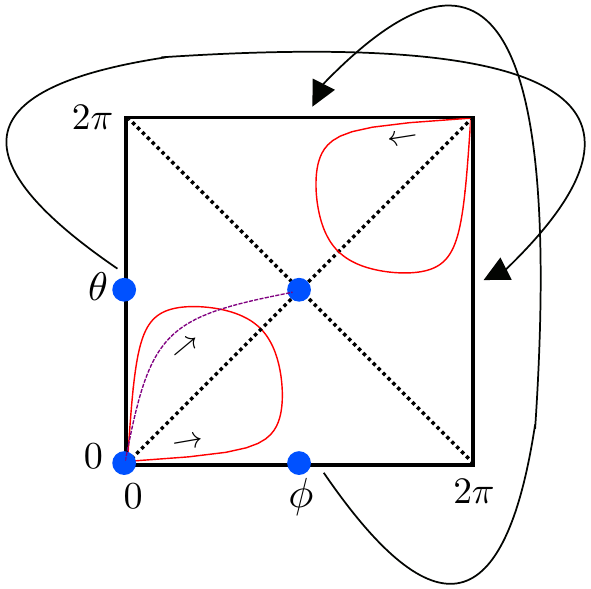}
\caption{The flat-torus manifold on which the phase shifts
  propagate. The blue dots correspond to the four fixed points of the
  RG. The dotted lines forming the diagonals of the square are axes of
  symmetry for $S$-matrix trajectories. The purple dashed curve is an example of an
  $S$-matrix trajectory in the scattering length approximation corresponding to the fourth
  row of Table~\ref{tab:LOsymclass} (with
  $a_1,a_0 >0$ and $a_1/a_0=5$). The red curves are examples of
  $S$-matrix trajectories with range parameter given in the fifth row
  of Table~\ref{tab:genconfrm}.  These trajectories leave
  $\phi+\theta$ invariant (with $a_1,a_0 <0$ (bottom left quadrant)
  and with $a_1,a_0 >0$ (top right quadrant), $a_0/a_1=15$ and
  $\lambda=0.01$).}
  \label{fig:ftbase}
\end{figure}

The $S$-matrix of Eq.~(\ref{eq:Sdefres}) is specified by the two
angular variables $\phi(p)$ and $\theta(p)$ that are determined by the
Schr\"odinger equation once the finite-range quantum mechanical
potential is specified. As these variables are periodic,
the two-dimensional ``phase space'' that these variables define is a
flat torus manifold, illustrated in Fig.~(\ref{fig:ftbase}). The range
of values that $\phi(p)$ and $\theta(p)$ can take are bounded by
unitarity, with boundary values determined by the four RG fixed points,
which occur at $\hat {\bf S} = \pm \hat {\bf 1}$ and $\pm \hat{{\cal
    P}}_{12}$ when the s-wave scattering lengths are either vanishing
or infinite (at unitarity)~\cite{Beane:2018oxh,Beane:2020wjl}.  Generally, in
effective range theory, the $S$-matrix trajectory on the flat torus
will originate at the trivial fixed point at scattering threshold and
trace out a curve that exits the flat-torus and enters a bulk space at
the first inelastic threshold~\cite{Beane:2020wjl}. In what follows,
all inelasticities will be pushed to infinite momentum and $S$-matrix
trajectories will begin and end at an RG fixed point.

\subsection{UV/IR symmetries of the $S$-matrix}

\subsubsection*{Out-state density matrix}

\noindent In this section $S$-matrices with a momentum inversion
symmetry that interchanges the IR and the UV will be studied. The $S$-matrix is
defined as the operator which evolves the incoming (``in'') state
before scattering into the outgoing (``out'') state after scattering,
i.e. $\hat {\bf S} \lvert \text{in} \rangle = \lvert \text{out}
\rangle$.  The criterion for the presence of a symmetry will be based on the
transformation properties of the density matrix of the ``out'' state
\begin{equation}
    \rho = \lvert \text{out} \rangle \langle \text{out} \rvert = \hat {\bf S} \lvert \text{in} \rangle \langle \text{in} \rvert \hat {\bf S}^{\dagger} \ .
\end{equation}
The first kind of symmetry transformation that will be considered is
$\rho \mapsto \rho$ which leaves all spin-observables invariant. In
the following sections it will be shown that there are non-trivial
instances of this symmetry where the $S$-matrix itself is not
invariant. The second kind of symmetry transformation that will be considered
is $\rho \mapsto \bar{\rho}$ where
\begin{equation}
    \bar{\rho} \equiv \hat {\bf S}^{*} \lvert \text{in} \rangle \langle \text{in} \rvert \hat {\bf S}^T \ .
\end{equation}
This is the density matrix that would be produced if all
phase shifts change sign, or, equivalently, if attractive
interactions are replaced with repulsive interactions of equal magnitude, and
vice-versa.

\subsubsection*{The scattering-length approximation}

\noindent The $S$-matrix takes constant values at the fixed points of
the RG, which implies that at both the trivial and unitary fixed
points, the underlying EFT has non-relativistic conformal invariance
(Sch\"odinger symmetry).  The scattering trajectories are bridges
which connect various conformal field theories. In addition to this
conformal invariance of the EFT action at the RG fixed points, there is a UV/IR symmetry
which acts directly on the $S$-matrix and which is present at finite
values of the scattering lengths. 

Consider the momentum inversion transformation
\begin{eqnarray}
p\mapsto \frac{1}{|a_1 a_0| p} \ .
   \label{eq:moebius}
\end{eqnarray}
As this maps $p=0$ to $p=\infty$ it is a transformation which interchanges the IR and UV.
It corresponds to the transformations on the phase shifts and density
matrices shown in Table~\ref{tab:LOsymclass}.  This momentum inversion
is a conformal invariance that leaves the combination of angular
variables $\phi+\theta$ ($a_1 a_0 <0$) or $\phi-\theta$ ($a_1 a_0 >0$)
invariant and implies a reflection symmetry of the $S$-matrix
trajectory, as illustrated for a specific case in
Fig.~(\ref{fig:ftbase}) (purple curve). When $a_1 a_0 > 0$, the two
phase shifts conspire to leave the density matrix
unchanged despite neither phase shift separately being invariant.
This demonstrates that, in multi-channel scattering, there
exist symmetries which are not manifest at the level of the scattering
amplitude and appear as reflections of $S$-matrix trajectories.  It
is notable that the EP is invariant with respect to the
transformation of Eq.~(\ref{eq:moebius})~\cite{Beane:2020wjl}.
\begin{table}[h!]
\centering
\begin{tabularx}{0.8\textwidth} { 
  | >{\raggedright\arraybackslash}X 
  | >{\centering\arraybackslash}X 
  | >{\centering\arraybackslash}X 
    | >{\centering\arraybackslash}X 
  | >{\raggedleft\arraybackslash}X | }
 \hline
  $\phi\mapsto$  & $\theta\mapsto$ & $\rho\mapsto$ & $a_0$ &  $a_1$  \\
\hlineB{3}
 $-{\pi} + \theta$  & ${\pi} + \phi$ & $\bar{\rho}$ & $+$ & $-$  \\
  \hline
${\pi} + \theta$  & $-{\pi} + \phi$ & $\bar{\rho}$ & $-$ & $+$ \\
\hline
${\pi} - \theta$  & ${\pi} - \phi$ & $\rho$ & $-$ & $-$ \\
 \hline
$-{\pi} - \theta$  & $-{\pi} - \phi$ & $\rho$ & $+$ & $+$ \\
\hlineB{2}
\end{tabularx}
  \caption{Action of the momentum inversion transformation in the scattering length approximation.}
 \label{tab:LOsymclass}
\end{table}

\subsubsection*{Including range corrections}

\noindent The scattering length approximation is a part of
a larger class of UV/IR symmetric $S$-matrix models which include range effects
that are strictly correlated with the scattering lengths. These
UV/IR symmetries have a distinct character as the range effects
necessarily arise from derivative operators in the EFT.
Consider the general momentum inversion
\begin{eqnarray}
p\mapsto \frac{1}{\lambda |a_1 a_0| p} \ ,
   \label{eq:moebiusgen}
\end{eqnarray}
where the real parameter $\lambda >0$. One can ask: what is the most
general $S$-matrix for which this inversion symmetry gives rise to
interesting symmetries? This transformation rules out all
shape-parameter effects and correlates the effective ranges with
the scattering lengths in a specific way. Table~\ref{tab:genconfrm}
gives the effective-range parameters for all $S$-matrix models with
a symmetry under the momentum inversion of Eq.~(\ref{eq:moebiusgen}).
Note that the first four rows correlate the singlet
effective range with the triplet scattering length and vice-versa.
\begin{table}[h!]
\centering
\begin{tabularx}{0.8\textwidth} { 
  | >{\raggedright\arraybackslash}X 
  | >{\centering\arraybackslash}X 
  | >{\centering\arraybackslash}X 
    | >{\centering\arraybackslash}X 
  | >{\raggedleft\arraybackslash}X | }
 \hline
  $\phi\mapsto$  & $\theta\mapsto$ & $\rho\mapsto$ & $r_0$ & $r_1$ \\
\hlineB{3}
 $\phi$  & $\theta$ & $\rho$ & $-2\eta/a_0$ & $-2\eta/a_1$ \\
\hline
$\phi$  & $-\theta$ & $\rho_-  + \bar{\rho}_+$ & $-2\eta/a_0$ & $+2\eta/a_1$ \\
 \hline
$-\phi$  & $\theta$ & $\rho_+ + \bar{\rho}_- $ & $+2\eta/a_0$ & $-2\eta/a_1$ \\
 \hline
$-\phi$  & $-\theta$ & $\bar{\rho}$ & $+2\eta/a_0$ & $+2\eta/a_1$ \\
\hlineB{2}
$\theta$  & $\phi$ & $\bar{\rho}$ & $-2\eta/a_1$ & $-2\eta/a_0$ \\
 \hline
$-\theta$  & $-\phi$ & $\rho$ & $+2\eta/a_1$ & $+2\eta/a_0$ \\
 \hline
\end{tabularx}
  \caption{Action of the momentum-inversion transformation on models that have a non-zero effective range. Here $\eta \equiv \lambda \lvert a_0 a_1 \rvert$ and $\rho_{\pm}$ is the density matrix projected onto the total spin $\frac{1}{2} \left ( 1 \pm 1 \right )$ subspace.}
 \label{tab:genconfrm}
\end{table}

An interesting and well-known feature of the effective-range expansion
in nucleon-nucleon (NN) scattering is the smallness of the shape parameter
corrections (see, for instance, Ref.~\cite{Cohen:1998jr,deSwart:1995ui}) as compared
to the range of the interaction, which is given roughly by the Compton
wavelength of the pion. Vanishing shape corrections is a key signature
of an $S$-matrix with momentum-inversion symmetry. Indeed, as the
NN s-wave effective ranges are positive while the
scattering lengths have opposite sign, the model given in the second
row of Table~\ref{tab:genconfrm}, with $\lambda$ fitted to the data,
provides a description of the low-energy s-wave phase shifts that
improves upon the scattering-length approximation. As will be seen
below, models that arise from zero-range forces with exact
momentum-inversion symmetry and a positive effective range strictly
violate causality. However, relaxing the zero-range condition
can lead to interesting results for nuclear physics, as 
is considered in Ref.~\cite{Beane:2021C}.

\section{Geometric scattering theory}
\label{sec:geom}

\subsection{Metric on the flat-torus}

\noindent As the space on which the two phase shifts, $\theta$ and $\phi$,  propagate is a two-dimensional flat space,
the line element should take the form $ds^2 \propto d\phi^2\;+\;  d\theta^2$. This metric
can be obtained formally by parameterizing the $S$-matrix of Eq.~(\ref{eq:Sdefres}) as
\begin{eqnarray}
  \hat {\bf S}   & = & \big\lbrack x(p)\;+\; i\;y(p) \big\rbrack    \hat   {\bf 1}
\ +\
\big\lbrack z(p)\;+\; i\;w(p) \big\rbrack \hat{{\cal P}}_{12} \ ,
   \label{eq:Sdefgen}
\end{eqnarray}
with
\begin{eqnarray}
  x & =&   \oneht\; \lbrack \cos(\phi) + \cos(\theta) \rbrack \ \ , \ \ 
  y \ =\   \oneht\; \lbrack \sin(\phi) + \sin(\theta) \rbrack\ \ , \ \ \nn \\
  z & =&   \oneht\; \lbrack -\cos(\phi) + \cos(\theta) \rbrack\ \ , \ \ 
  w \ =\   \oneht\; \lbrack -\sin(\phi) + \sin(\theta) \rbrack \ .
   \label{eq:PS}
\end{eqnarray}
Then, as an embedding in $\mathbb{R}^4$, with line
element
\begin{eqnarray}\;
ds^2 \ =\ dx^2\;+\;dy^2\;+\;dz^2\;+\;dw^2 \ ,
   \label{eq:R4metricgen}
\end{eqnarray}
one finds the flat two-dimensional Euclidean line element
\begin{eqnarray}\;
ds^2 \ =\ \oneht\;\left( d\phi^2\;+\;  d\theta^2\right)\ .
   \label{eq:T2inR4}
\end{eqnarray}
With $\phi$ and $\theta$ periodic, the corresponding metric describes the flat torus $\mathbb{T}^2\sim S^1 \times S^1\hookrightarrow \mathbb{R}^4$, where $S^1$
is the circle. From this line element, one can read off the flat-torus metric tensor $g_{ab}$, with $a,b=1,2$.

\subsection{Geometric action} \label{sec:geoact}

\noindent The action for a general parameterization of a curve on a space with
coordinates $\mathcal{X}^1=\phi$ and $\mathcal{X}^2=\theta$ and metric tensor $g_{ab}$
can be expressed as~\cite{garay2019classical}
 \begin{equation}
 \int L\left(\mathcal{X},\dot{\mathcal{X}} \right)d\sigma \ =\ \int \left({\bf{N}}^{-2}g_{ab} \dot{\mathcal{X}}^a \dot{\mathcal{X}}^b \ -\  \mathbb{V(\mathcal{X})} \right){\bf{N}}d\sigma \ ,
    \label{eq:action}
 \end{equation}
 where $\sigma$ parameterizes the curve (affine or inaffine), $L$ is the Lagrangian,
 $\dot{\mathcal{X}}\equiv{d\mathcal{X}}/{d\sigma}$, and
 $\mathbb{V(\mathcal{X})}$ is an external geometric potential which is assumed
 to be a function of $\mathcal{X}$ only. The corresponding
 Euler-Lagrange equations give the trajectory equations
 \begin{equation}
\ddot{\mathcal{X}}^a \ +\ {}_g\Gamma^a_{\ bc} \dot{\mathcal{X}}^b \dot{\mathcal{X}}^c \ =\ \kappa(\sigma) \dot{\mathcal{X}}^a
   \ -\ \oneht {\bf{N}}^2 g^{ab}  \partial_b\mathbb{V(\mathcal{X})} \ ,
    \label{eq:exEOMfromact}
 \end{equation}
 where ${}_g\Gamma^a_{\ bc}$ are the Christoffel symbols for the metric $g_{ab}$, and
 \begin{equation}
\kappa(\sigma) \ \equiv \ \frac{\dot{\bf{N}}}{\bf{N}} = \frac{d}{d\sigma}\ln \frac{d\tau}{d\sigma} \ .
    \label{eq:inaffinity}
 \end{equation}
 Here $\kappa$ is the inaffinity~\cite{blau2020}, which vanishes when $\sigma=\tau$ with $\tau$ an affine parameter. 
 
 An interesting feature of the geometric construction of scattering is that the relative momentum
 that describes the motion of the center-of-mass is a non-affine parameter. For a constant geometric potential, the trajectory equations reduce to the geodesic equations
 which describe straight-line trajectories on the flat-torus. Any curvature indicates the presence of a non-constant geometric potential. Now, {\it a priori}, if a solution for $\phi$ and $\theta$ is specified,
 there are two equations of motion for three unknowns, the inaffinity and the two force components in the $\phi$ and $\theta$ directions. However, the presence of UV/IR symmetries can reduce the number of
 unknowns to two and thus allows an explicit construction of the geometric potential~\cite{Beane:2020wjl}.

\subsection{Solvable models}

\noindent In the scattering length approximation, the UV/IR symmetry of Eq.~(\ref{eq:moebius}) determines the geometric potential exactly. It is given by~\cite{Beane:2020wjl}
\begin{eqnarray}
\mathbb{V}(\phi,\theta) & = &  \, \frac{|a_0 a_1|}{\left(|a_0|+|a_1| \right)^2c_1^2} \tan^2\left(\oneht(\phi+\epsilon\,\theta)\right) \ ,
\label{eq:entPOT} 
\end{eqnarray}
where $\epsilon\,=-1$ for $a_1 a_0 >0$ and $\epsilon\,=+1$ for $a_1 a_0 <0$, and $c_1$ is an
integration constant. The inaffinity associated with a trajectory parameterized by the c.o.m momentum is constructed from
\begin{eqnarray}
{\bf{N}} &=& \frac{c_1}{p}\left(\sin\phi -\epsilon \sin\theta \right) \ .
\label{eq:Nfun} 
\end{eqnarray}
The $S$-matrix trajectory is independent of the parameterization and can be simply described
---with vanishing inaffinity and choice ${\bf{N}}={c_1}=1$--- by an affine parameter $\tau$ 
via the simple Lagrangian
\begin{eqnarray}
L &=& \oneht \left( \dot{\phi}\,+\,\dot{\theta} \right) \ -\ \mathbb{V}(\phi,\theta) \ ,
\label{eq:LagAFFLO}
\end{eqnarray}
where the dots denote differentiation with respect to $\tau$. Of course $\tau$ has no interpretation as a momentum or energy in a scattering
process; such a parameter is not affine.

The conformal $S$-matrix models with the UV/IR symmetry of Eq.~(\ref{eq:moebiusgen}), and $\lambda$-dependent effective ranges, also lead to solvable geometric
potentials. The general solution is cumbersome, however in the special case where the effective ranges
are correlated to the scattering lengths as in the last two rows of Table~\ref{tab:genconfrm} ---and $\lambda=1/4$--- the geometric potential is identical to Eq.~(\ref{eq:entPOT}) except for an overall factor of $1/2$ and a rescaled argument
\begin{eqnarray}
\oneht(\phi+\epsilon\,\theta) & \longrightarrow & \onefourth(\phi+\epsilon\,\theta) \ .
\label{eq:entPOTlam14} 
\end{eqnarray}
It will be seen in section~\ref{sec:causalsing} why this case is special.

\section{Spacetime constraints}
\label{sec:stcon}

\subsection{Galilean invariance}

\noindent The parameter $p$, which labels the c.o.m. momentum of the scattering process,
is related to the total energy, $E$, in the system by $p =
\sqrt{M E}$. If the incoming particle momenta are ${\vec p}_1$ and
${\vec p}_2$ then in the c.o.m. frame, ${\vec p}_1 = - {\vec p}_2
\equiv \vec{\underline{p}}$ and $p = \lvert \, \vec{\underline{p}} \,
\rvert$. Other Galilean frames can be reached from the c.o.m. frame
via a combined rotation ${\bf{\cal R}}$ and boost by a velocity ${\vec
  v}$:
\begin{eqnarray}
  \vec{\underline{p}} \ && \longrightarrow \ {\bf{\cal R}} \, \vec{\underline{p}} + M {\vec v}
 \label{eq:Galilean}
\end{eqnarray}
which implies the transformation
\begin{eqnarray}
  p \ && \longrightarrow \  \lvert \, \vec{\underline{p}} \, \rvert \, \sqrt{1+{\bf x}} \ ,
 \label{eq:Galilean2}
\end{eqnarray}
with ${\bf x}\equiv {(M {\vec v\,})^2}/{{\vec {\underline{p}}}^2}$. Varying ${\bf x}$ between $0$ and $1$ corresponds to transforming
between the c.o.m. and laboratory frames. Hence, Galilean invariance allows arbitrary reparameterizations of the $S$-matrix
of the form $p \rightarrow \Omega \, p$ with $1\leq \Omega \leq\infty$ and $\Omega$ interpolating between the c.o.m at rest
and boosted to infinite momentum. As this is just a rescaling of $p$, changing to another inertial frame does not affect the inaffinity, Eq.~(\ref{eq:inaffinity}).

\subsection{Causality bounds on zero-range scattering}

\subsubsection*{Wigner bounds}

\begin{figure}[!ht]
\centering
\includegraphics[width = 0.55\textwidth]{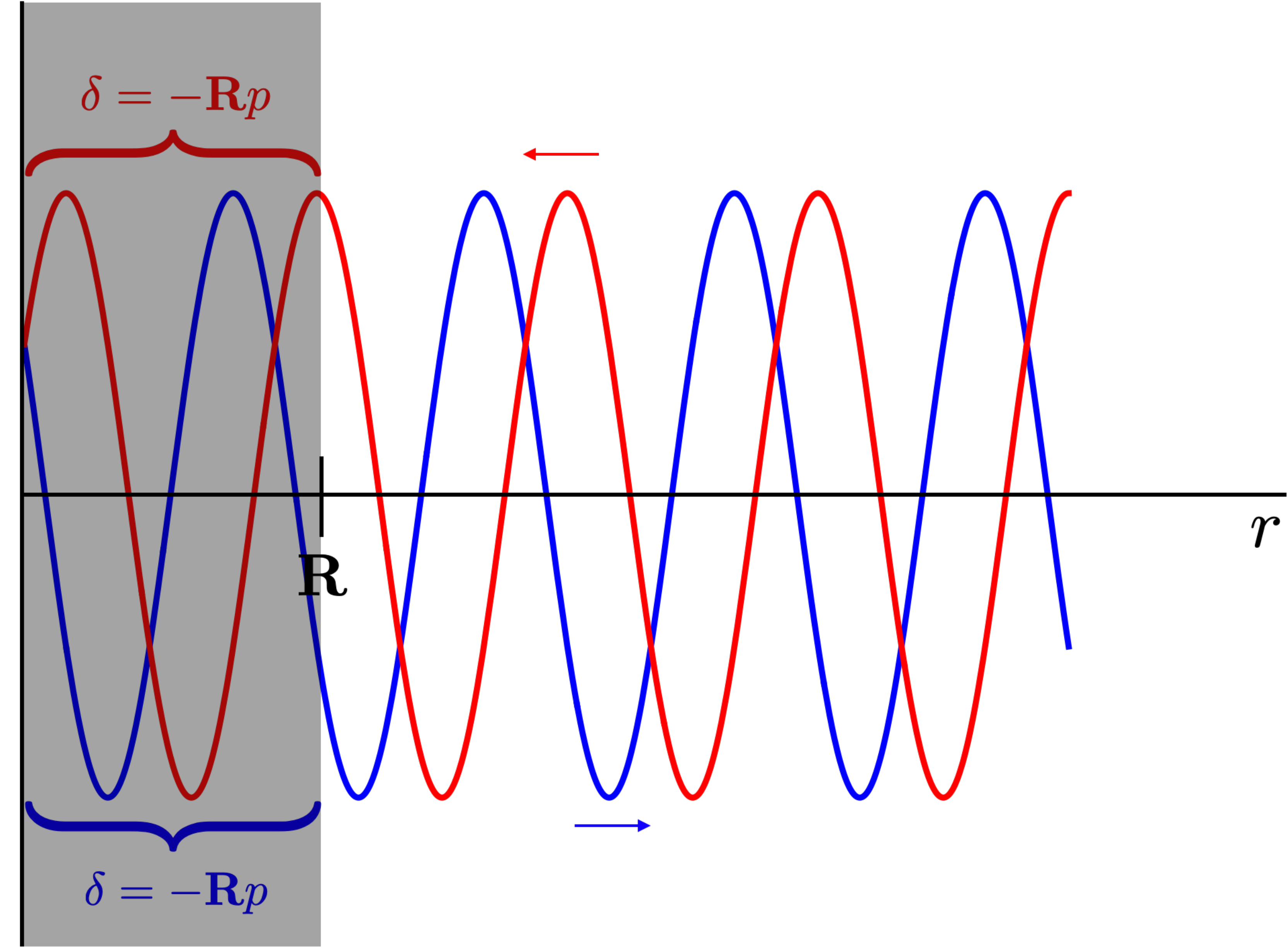}
\caption{An illustration of the minimum phase shift for scattering off a potential of range ${\bf R}$. The red (blue) curve represents the phase of the incoming (outgoing) s-wave spherical wave.}
  \label{fig:ClassWigBound}
\end{figure}
\noindent In non-relativistic scattering with finite-range forces, causality places bounds on
physical scattering parameters by way of Wigner
bounds~\cite{Wigner:1955zz,Phillips:1996ae,Hammer:2009zh,Hammer:2010fw}.
Consider a two-body s-wave wave function 
both free and in the presence of an interaction potential of range
${\bf R}$. The difference in phase between the scattered and free spherical wave is
defined to be twice the phase shift. The most negative phase shift is
obtained when the scattered wave does not penetrate the potential 
and reflects off the boundary at $r = {\bf R}$. This provides a
lower bound on the phase shift $\delta(p) \geq - {\bf R} p$, see
Fig.~(\ref{fig:ClassWigBound}). Now consider a plane wave at an
infinitesimally larger momentum, $\bar{p}$ with $\delta(\bar{p}) \geq -
{\bf R} \bar{p} $. The difference between $\delta(\bar{p})$ and $\delta(p)$
provides a semi-classical bound on the derivative of the phase shift with respect to
momentum
\begin{equation}
    \frac{d \delta}{dp} \geq - {\bf R} \ .
  \label{eq:WBclass}
\end{equation}
By time evolving the plane waves, the above becomes a bound on the time delay between the incident and scattered wave, $\Delta t \geq -{M {\bf R}}/{p}$. It is in this sense that causality constrains non-relativistic scattering.
A more careful derivation, which includes quantum mechanical effects,
induces a second term on the right hand side of Eq.~(\ref{eq:WBclass})
and leads to the bound~\cite{Wigner:1955zz}:
\begin{eqnarray}
  \frac{d \delta}{dp} \geq -{\bf R} \ +\ \frac{\sin \left(2\delta+2 p {\bf R}\right)}{2p}  \ .
  \label{eq:WB3dbp1}
\end{eqnarray}
Evaluated at threshold this becomes a constraint on the effective range parameter:
\begin{eqnarray}
r\ \leq \ 2\bigg\lbrack {\bf R}\, -\, \frac{{\bf R}^2}{a} \, +\, \frac{{\bf R}^3}{3 a^2} \bigg\rbrack \ .
  \label{eq:WB3da}
\end{eqnarray}

In the Wilsonian EFT paradigm, an $S$-matrix element derived from EFT
is dependent on a momentum cutoff, $\Lambda\sim 1/{\bf R}$, which is
kept finite and varied to ensure cutoff-independence to a given order
in the perturbative EFT expansion.  What occurs above this scale is
irrelevant to the infrared physics that is encoded by the
$S$-matrix and compared to experiment. An explicit calculation of the
Wigner bound in the EFT of contact operators with cutoff
regularization can be found in Ref.~\cite{Phillips:1996ae}.  As the
bound depends explicitly on the EFT cutoff, its relevance in
physical scenarios is somewhat ambiguous as the EFT can violate
causality bounds as long as the violations occur above the cutoff,
and the bound itself weakens as higher-order corrections in
the EFT expansion are included~\cite{Beck:2019abp}.
\begin{table}[h!]
\centering
\begin{tabularx}{0.9\textwidth} { 
  | >{\raggedright\arraybackslash}X 
  | >{\centering\arraybackslash}X 
  | >{\centering\arraybackslash}X 
    | >{\centering\arraybackslash}X 
  | >{\raggedleft\arraybackslash}X | }
 \hline
  $\phi\mapsto$  & $\theta\mapsto$  & $\rho \mapsto$ & $r_0$ & $r_1$ \\
\hlineB{3}
 $\phi$  & $\theta$ & $\rho$ &   $-2a_1\lambda$  \ \ ${\scriptstyle (a_1>0)}$ & $-2a_0\lambda$ \ \  ${\scriptstyle (a_0>0)}$\\
\hline
$\phi$  & $-\theta$ & $ \rho_- +  \bar{\rho}_+$ & $+2a_1\lambda$ \ \ ${\scriptstyle (a_1<0)}$& $-2a_0\lambda$ \ \ ${\scriptstyle (a_0>0)}$\\
 \hline
$-\phi$  & $\theta$ & $ \rho_+ +  \bar{\rho}_-$ & $-2a_1\lambda$ \ \ ${\scriptstyle (a_1>0)}$& $+2a_0\lambda$ \ \ ${\scriptstyle (a_0<0)}$\\
 \hline
$-\phi$  & $-\theta$ & $\bar{\rho}$ & $+2a_1\lambda$ \ \ ${\scriptstyle (a_1<0)}$& $+2a_0\lambda$ \ \ ${\scriptstyle (a_0<0)}$\\
\hlineB{2} 
$\theta$  & $\phi$ & $\bar{\rho}$ & $-2a_0\lambda$ \ \ ${\scriptstyle (a_0>0)}$& $-2a_1\lambda$ \ \ ${\scriptstyle (a_1>0)}$\\
 \hline
$-\theta$  & $-\phi$ & $\rho$& $+2a_0\lambda$ \ \ ${\scriptstyle (a_0<0)}$& $+2a_1\lambda$ \ \ ${\scriptstyle (a_1<0)}$\\
 \hline
\end{tabularx}
  \caption{Action of the momentum-inversion transformation on causal models that have a non-zero effective range.}
 \label{tab:genconfrmwb}
\end{table}

\begin{figure}[!ht]
\centering
\includegraphics[width = 0.8\textwidth]{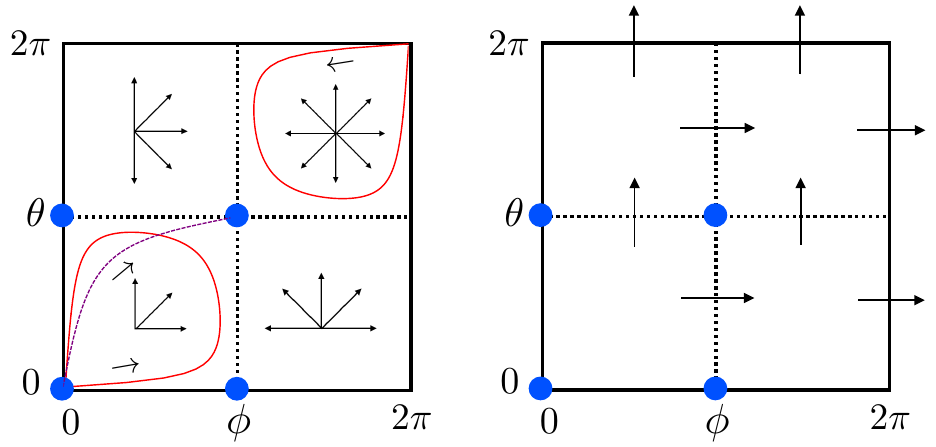}
\caption{Left panel: the range of tangent vectors on the flat
  torus allowed by the Wigner bound is superposed on Fig.~(\ref{fig:ftbase}).
  The purple trajectory in the bottom left quadrant has zero effective range
  and is seen to be consistent with the tangent-vector conditions.
  The red trajectory in the bottom left quadrant has positive effective ranges
  and is seen to violate the tangent-vector conditions, while the trajectory
  in the top right quadrant has negative effective ranges and is consistent
  with the conditions. Right panel: by matching the allowed tangent
  vectors at the boundaries of each quadrant it is found that
  $S$-matrix trajectories can only exit a quadrant via the upper or
  right edge.}
  \label{fig:WigTan}
\end{figure}

The $S$-matrix models with momentum-inversion symmetry can originate from
zero-range or finite-range forces. Here it will be assumed that the
underlying theory has strictly zero-range forces. This then implies
strong causality bounds whose geometric interpretation can be studied.
Explicitly, with zero-range forces, causality requires $r_s \leq 0$ and
the tangent vectors to $S$-matrix trajectories satisfy
\begin{eqnarray}
\dot{\phi}(p) \geq \frac{\sin\phi(p)}{p} \  \ \ , \ \ \ \dot{\theta}(p) \geq \frac{\sin\theta(p)}{p} \ ,
  \label{eq:WB3dx}
\end{eqnarray}
where dot represents differentiation with respect to momenta.  The
allowed tangent vectors clearly depend on the quadrant of the flat
torus in which they lie. In addition, by enforcing continuity of the
tangent vectors at the boundary of each quadrant, it is found that an
$S$-matrix trajectory can only exit a quadrant through the upper or
right edge. These various geometric constraints are illustrated in
Fig.~(\ref{fig:WigTan}) using the examples of Fig.~(\ref{fig:ftbase}).
It is notable that the large momentum behavior of any $S$-matrix curve
which ends at the trivial fixed point must be in the top-right
quadrant, which is also the only place where a trajectory can have
loops. Since the Wigner bound segregates by quadrant and indicates a
direction of preferred $S$-matrix evolution, causality introduces an
asymmetry which breaks the homogeneity and discrete isotropy of a
generic flat torus.
Applying the Wigner bound to the symmetric $S$-matrix models in
Table~\ref{tab:genconfrm} restricts the allowed signs of the
scattering lengths as shown in Table~\ref{tab:genconfrmwb}.

\subsubsection*{Causal singularities of the $S$-matrix}
\label{sec:causalsing}

\noindent In addition to Wigner bounds, causality in non-relativistic scattering is manifest in various constraints on the analytic structure
of the $S$-matrix in the complex-momentum plane~\cite{PhysRev.74.131,PhysRev.83.249,PhysRev.91.1267}. The simplicity of the $S$-matrix
models with momentum-inversion symmetry reveal these constraints and their relation with the Wigner bound in straightforward fashion.
The s-wave $S$-matrix elements with momentum inversion symmetry are ratios of polynomials of second degree and can thus be expressed as
\begin{eqnarray}
S_s &\equiv&\frac{\left(p+p_s^{\sss(1)}\right)\left(p+p_s^{\sss(2)}\right)}{\left(p-p_s^{\sss(1)}\right)\left(p-p_s^{\sss(2)}\right)} \ ,
   \label{eq:Selementres}
\end{eqnarray}
where
\begin{eqnarray}
p_s^{\sss(1,2)}&=& \frac{1}{r_s}\left(i\pm \sqrt{\frac{2r_s}{a_s}-1}\right) \ .
   \label{eq:polposres}
\end{eqnarray}
Consider the evolution of the singularities in the complex-$p$ plane as $\lambda$ is varied~\cite{Habashi:2020qgw}
for the causal model given in the last row of Table~\ref{tab:genconfrmwb}.
This model, with both scattering lengths negative, leaves $\phi-\theta$ invariant, and
has poles at
\begin{eqnarray}
p_s^{\sss(1,2)}&=& -\frac{1}{2|a_s|\lambda}\left(i\pm \sqrt{4\lambda-1}\right) \ .
   \label{eq:polposresmod}
\end{eqnarray}

\begin{figure}[!ht]
\centering
\includegraphics[width = 0.65\textwidth]{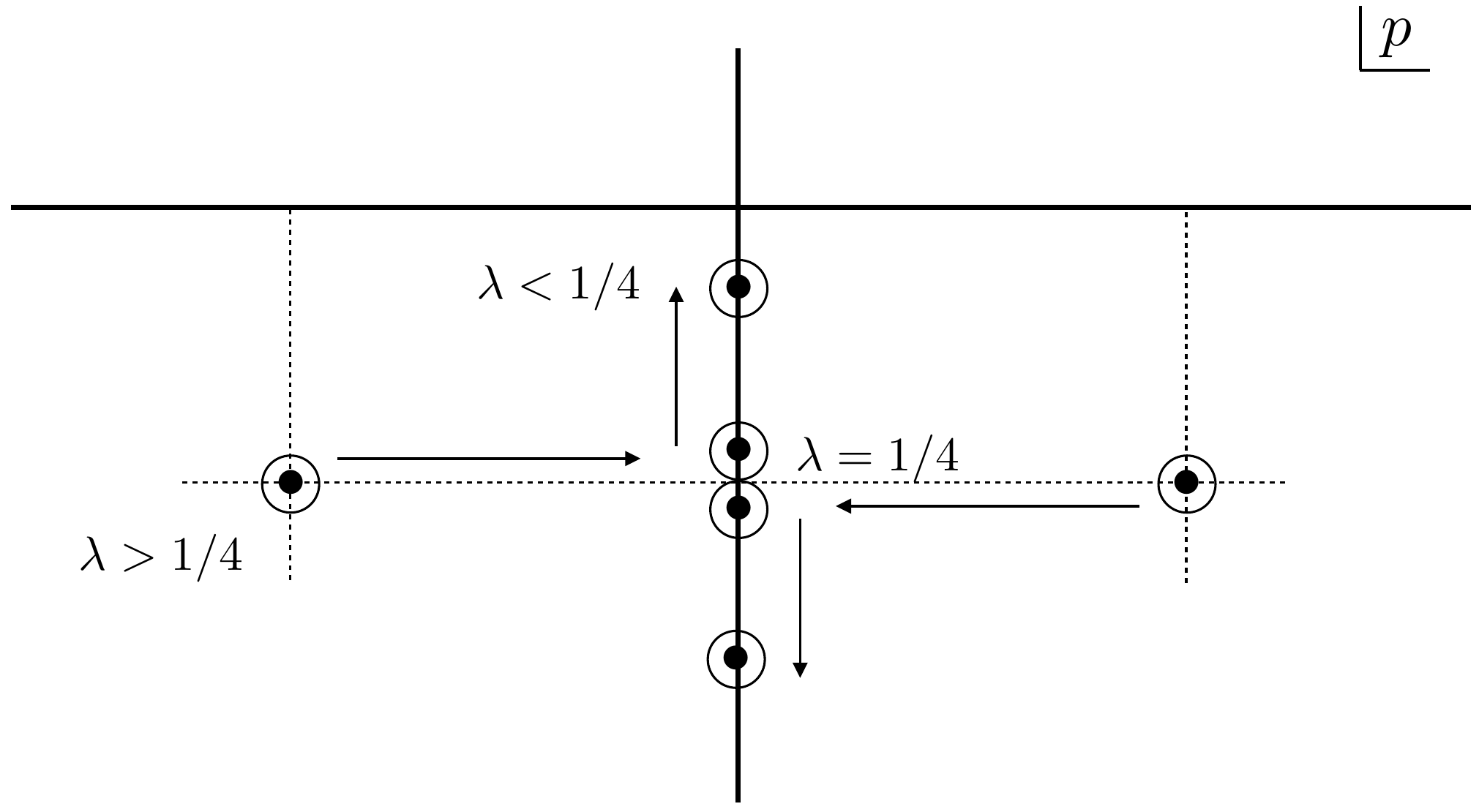}
\caption{Singularities of $S$-matrix elements in the complex-$p$ plane as $\lambda$ is varied. The arrows indicate direction of decreasing $\lambda$.}
  \label{fig:poles}
\end{figure}

There are three distinct cases, illustrated in Fig.~(\ref{fig:poles}).

\noindent $\underline{\lambda > 1/4}$: there are two resonance poles in the lower-half complex plane on
opposite sides of the imaginary axis. Dropping the partial-wave subscript on the scattering length,
\begin{eqnarray}
  p^{\sss(1)}&=& -p_R-i p_I \ \ \ , \ \ \
  p^{\sss(2)}= p_R-i p_I \ ,
   \label{eq:polposcase1a}
\end{eqnarray}
with
\begin{eqnarray}
  p_R &=& \frac{1}{2|a|\lambda}\sqrt{4\lambda-1}\ \ , \ \ p_I = \frac{1}{2|a|\lambda} \ ;
  \label{eq:polposcase1b}
\end{eqnarray}
$\underline{\lambda = 1/4}$: there is a double pole corresponding to a virtual state on the negative imaginary axis at
\begin{eqnarray}
  p^{\sss(1)} &=&  p^{\sss(2)}= -\frac{i}{2|a|\lambda} \ ;
   \label{eq:polposcase2}
\end{eqnarray}
$\underline{\lambda < 1/4}$: there are two poles corresponding to virtual states on the negative imaginary axis at
\begin{eqnarray}
  p^{\sss(1)}&=& -i p_- \ \ \ , \ \ \
  p^{\sss(2)}= -i p_+ \ ,
   \label{eq:polposcase3a}
\end{eqnarray}
with
\begin{eqnarray}
  p_\pm &=& \frac{1}{2|a_s|\lambda}\left(1\pm \sqrt{1-4\lambda}\right) \ .
  \label{eq:polposcase3b}
\end{eqnarray}
It is clear that the Wigner bound implies that the poles of the $S$-matrix elements lie in the lower-half of the complex-momentum plane as
one would expect of states that decay with time. Note that the special case of the double pole (and vanishing square root)
is in correspondence with the special case geometric potential which takes the simple form, Eq.~(\ref{eq:entPOTlam14}).

\subsection{Spatial dependence of scattering}
\label{subsec:2dscat}

\noindent The $S$-matrix is clearly aware of the number of spatial
dimensions that it is acting in and one therefore expects that this is
reflected in the geometric theory of scattering via a modified
geometric potential. It is straightforward to carry out the analysis
that was done above in two spatial dimensions. In two dimensions, low energy
scattering arising from short-range forces is enhanced in the IR due to an apparent scaling symmetry of the Schr\"odinger equation. One consequence of this is that there is only a single fixed point $S$-matrix, the identity, which is reached at both zero and
infinite scattering length. Spin and particle statistics are also distinct in
two dimensions, however, for our purposes, all that will be required is a
scattering process with two independent low-energy channels. One way this could be achieved is by placing the three-dimensional scattering system in a strongly
anisotropic harmonic potential which effectively confines one of the
spatial dimensions~\cite{PhysRevA.64.012706,PhysRevA.76.063610,PhysRevA.85.061604,PhysRevA.98.051603,PhysRevLett.94.210401}. This allows
the two-fermion system to be continuously deformed from three to two
dimensions and provides a means of studying the dependence of
the geometric theory, constructed above, on spatial dimensionality.
A qualitatively equivalent and simpler way of achieving this reduction of
dimensionality is by periodically identifying and compactifying one of
the spatial dimensions~\cite{PhysRevA.93.063631,Beane:2018huc}, say in the
$z$-direction. 

Regardless of how the two dimensional system is obtained, the ERE is~\cite{Braaten:2004rn,Beane:2010ny,Kaplan:2005es}
\begin{eqnarray}
\cot{\delta_s(p)} & =& {1 \over \pi} \log({\textit{\textbf a}}^2_s p^2) + \sigma_{2, \, s} p^2 + \mathcal{O}(p^4)
 \label{eq:2DERE}
\end{eqnarray}
where the ${\textit{\textbf a}}_s$ and $\sigma_{2, \, s}$ are the two-dimensional scattering lengths and areas, respectively. The full $S$-matrix can be constructed from the phase shifts as in Eq.~(\ref{eq:Sdef}). Retaining just the first term in the ERE gives rise to the scattering length approximation which, in terms of periodic variables on the flat torus, is
\begin{eqnarray}
\phi & =& 2\cot^{-1}\!\left({1 \over \pi} \log({\textit{\textbf a}}^2_0 p^2)\right) \ \ \ , \ \ \ \theta\ =\ 2\cot^{-1}\!\left({1 \over \pi}\log({\textit{\textbf a}}_1^2 p^2)\right) \ ,
 \label{eq:LOinaffineSOL2D}
\end{eqnarray}
where the higher order effective area and shape parameters have
been set to zero\footnote{This can be obtained from the
compactification of a spatial dimension if the $d=3$ effective range
parameters are functions of the compactification
radius~\cite{Beane:2018huc}.}. Note that there is an IR enhancement in two dimensions, as made evident by the logarithmic dependence on the c.o.m. momenta, and that there exists a bound state for either sign of coupling strength~\cite{Beane:2010ny,Kaplan:2005es,Beane:2021A}.
  
The momentum inversion transformation takes a similar form as in three dimensions,
$p\to ({\textit{\textbf a}}_1 {\textit{\textbf a}}_0\, p)^{-1}$, and the
phase shifts transform as
\begin{eqnarray} 
   \phi(p) \mapsto -\theta(p) \ \ & ,& \ \ \theta(p) \mapsto -\phi(p) \ ,
  \label{eq:confmoebiusiso22d}
\end{eqnarray}
which leaves the density matrix invariant.
In two spatial dimensions, the momentum-inversion symmetry implies that all effective area and
shape parameters must vanish\footnote{In $d=2$, causality bounds the effective area
parameter~\cite{Hammer:2009zh,Hammer:2010fw}
\begin{eqnarray}
\sigma_{2,\,s}\ \leq \ \frac{{\bf R}^2}{\pi} \bigg\lbrace \bigg\lbrack \log\left(\frac{\bf R}{2{\textit{\textbf a}}_s}\right)\, +\, {\gamma} \, -\, \frac{1}{2} \bigg\rbrack^2  \,+\,\frac{1}{4} \bigg\rbrace \ ,
  \label{eq:WB2da}
\end{eqnarray}
where $\gamma$ is the Euler-Mascheroni constant. Therefore, momentum-inversion implies that the Wigner bound is saturated with $\sigma_{2,\,s} = 0$.}.
The geometric potential on the flat torus which reproduces the phase shifts of Eq.~(\ref{eq:LOinaffineSOL2D}) is found to be
\begin{eqnarray}
\mathbb{V}(\phi,\theta) & = &  \, -\frac{\pi^2}{4\left(\log({{\textit{\textbf a}}_0}/{{\textit{\textbf a}}_1})\right)^2c_1^2} \tan^2\left(\oneht(\phi+\theta)+\frac{\pi}{2}\right) \ .
\label{eq:entPOT2d} 
\end{eqnarray}
Notice that the harmonic dependence is the same as in three spatial
dimensions, Eq.~(\ref{eq:entPOT}), except for an additional phase of
$\pi/2$ which causes the geometric potential to diverge when both phase shifts
sum to zero. This can only occur at threshold, and can be attributed
to the infinite force needed to reproduce the singular behavior of the
phase shift derivatives at $p=0$.  Another property of the
geometric potential is the divergence of the prefactor when ${\textit{\textbf
    a}}_0 = {\textit{\textbf a}}_1$. At the end of
section~(\ref{sec:geoact}) it was pointed out that, for UV/IR symmetric
trajectories, there are two trajectory equations for two unknowns, the
inaffinity and the geometric potential. However, when the scattering
lengths are equal, and $\phi = \theta$, the two equations are no
longer linearly independent. In this case the trajectory is a geodesic
---a straight line--- on the flat torus, and no geometric potential is needed.

\section{Conclusion}
\label{sec:conc}

\noindent The $S$-matrix is a unitary operator that
evolves a state vector from the boundary of spacetime, into the
spacetime bulk to experience interaction, and then back to the
spacetime boundary.  In this view of scattering, all spacetime
features like causality and spatial dimensionality are bulk
properties, and, as the $S$-matrix is purely a function of kinematical
variables like momentum and energy, the bulk properties must be
imprinted in some way on these variables. In a general scattering
process, the $S$-matrix evolves an initial unentangled product state
into an entangled state which in general experiences non-local correlations.
In order to avoid the assumption of locality, which is intrinsic to
the EFT paradigm, Ref.~\cite{Beane:2020wjl} formulated a geometric
theory of scattering for two species of spin-1/2 fermions interacting
at low-energies via finite-range interactions. In this geometric theory 
the $S$-matrix emerges, without direct reference to spacetime, as a trajectory 
in an abstract space that is defined by unitarity. These $S$-matrix trajectories 
are generated by an entangling harmonic force whose form is 
---in certain special cases--- determined exactly by a UV/IR symmetry. 
This paper has considered the manner
in which causality and spatial dimensionality manifest themselves in
the geometric picture.

\vskip0.1in The main results of this paper are:

\vskip0.2in
\noindent$\bullet$ In $S$-matrix models with momentum-inversion
symmetry which arise from zero-range forces, Wigner bounds, which
enforce causality, severely restrict the model space by requiring that
effective ranges be strictly negative. These causality bounds place
restrictions on the tangent vectors of $S$-matrix trajectories which
propagate on the flat-torus in the geometric theory of scattering. This
preferred direction in the geometric space is the manifestation of
spacetime causality.

\vskip0.2in
\noindent$\bullet$ The geometric potential which determines the
$S$-matrix in solvable models with UV/IR symmetry takes the same
functional form in two and three spatial dimensions. The dependence on
spatial dimensionality is manifest both in the coupling strength and
phase of the geometric potential. This suggests that requiring unitarity and an ERE ensures that there are aspects of low-energy scattering, captured by the geometric formulation, that are universal and independent of spatial geometry. 
This could be anticipated given that the geometric potential leads to an entangling force, and the quantum correlations arising
from spin entanglement are a property of an internal Hilbert space.

\vskip0.2in
\noindent$\bullet$ The density matrix of the ``out'' state has been found to be the relevant object for identifying symmetries of a multi-channel scattering processes. All UV/IR symmetries in low-energy non-relativistic scattering with finite-range forces generated by momentum inversion that leave the density matrix invariant have been categorized and can be found in Tables~\ref{tab:LOsymclass} and \ref{tab:genconfrm}.

\vskip0.2in This work suggests many related investigations; among them
are: implications of the UV/IR symmetries of the $S$-matrix for the
construction of EFTs of s-wave NN scattering (this is considered in
Ref.~\cite{Beane:2021C}), as well as for EFT descriptions of few-
and many-body systems of nucleons.  Finally, the EPs of the $\pi\pi$
and $\pi$N $S$-matrices were recently considered in Ref.~\cite{Beane:2021zvo} and it
would be of interest to explore whether these scattering systems possess
analogous geometric constructions.


\section*{Acknowledgments}

\noindent We would like to thank David B.~Kaplan
for valuable discussions regarding this work. This work was supported by the U.~S.~Department of Energy
grants {\bf DE-FG02-97ER-41014} (UW Nuclear Theory) and {\bf DE-SC0020970}
(InQubator for Quantum Simulation).

\bibliographystyle{JHEP}
\bibliography{bibi}

\providecommand{\href}[2]{#2}\begingroup\raggedright\begin{thebibliography}{10}

\bibitem{Beane:2020wjl}
S.R.~Beane and R.C.~Farrell, \emph{{Geometry and entanglement in the scattering
  matrix}},
  \href{https://doi.org/https://doi.org/10.1016/j.aop.2021.168581}{\emph{Annals
  of Physics} {\bfseries 433} (2021) 168581}
  [\href{https://arxiv.org/abs/2011.01278}{{\ttfamily 2011.01278}}].

\bibitem{Arkani-Hamed:2020gyp}
N.~Arkani-Hamed, M.~Pate, A.-M.~Raclariu and A.~Strominger, \emph{{Celestial
  amplitudes from UV to IR}},
  \href{https://doi.org/10.1007/JHEP08(2021)062}{\emph{JHEP} {\bfseries 08}
  (2021) 062} [\href{https://arxiv.org/abs/2012.04208}{{\ttfamily
  2012.04208}}].

\bibitem{Arkani-Hamed:2019vag}
N.~Arkani-Hamed, S.~He, G.~Salvatori and H.~Thomas, \emph{{Causal Diamonds,
  Cluster Polytopes and Scattering Amplitudes}},
  \href{https://arxiv.org/abs/1912.12948}{{\ttfamily 1912.12948}}.

\bibitem{Beane:2000fx}
S.R.~Beane, P.F.~Bedaque, W.C.~Haxton, D.R.~Phillips and M.J.~Savage,
  \emph{{From hadrons to nuclei: Crossing the border}},
  \href{https://arxiv.org/abs/nucl-th/0008064}{{\ttfamily nucl-th/0008064}}.

\bibitem{Hammer:2019poc}
H.W.~Hammer, S.~K\"onig and U.~van Kolck, \emph{{Nuclear effective field
  theory: status and perspectives}},
  \href{https://doi.org/10.1103/RevModPhys.92.025004}{\emph{Rev. Mod. Phys.}
  {\bfseries 92} (2020) 025004}
  [\href{https://arxiv.org/abs/1906.12122}{{\ttfamily 1906.12122}}].

\bibitem{Beane:2021A}
S.R.~Beane and R.C.~Farrell, \emph{{UV/IR symmetries of the $S$-matrix and RG
  flow}},  \href{https://arxiv.org/abs/2112.03472}{{\ttfamily 2112.03472}}.

\bibitem{PhysRev.74.131}
N.~Hu, \emph{On the application of {H}eisenberg's theory of ${S}$-matrix to the
  problems of resonance scattering and reactions in nuclear physics},
  \href{https://doi.org/10.1103/PhysRev.74.131}{\emph{Phys. Rev.} {\bfseries
  74} (1948) 131}.

\bibitem{PhysRev.83.249}
W.~Sch\"utzer and J.~Tiomno, \emph{On the connection of the scattering and
  derivative matrices with causality},
  \href{https://doi.org/10.1103/PhysRev.83.249}{\emph{Phys. Rev.} {\bfseries
  83} (1951) 249}.

\bibitem{PhysRev.91.1267}
N.G.~van Kampen, \emph{${S}$- matrix and causality condition. {II.}
  nonrelativistic particles},
  \href{https://doi.org/10.1103/PhysRev.91.1267}{\emph{Phys. Rev.} {\bfseries
  91} (1953) 1267}.

\bibitem{Wigner:1955zz}
E.P.~Wigner, \emph{{Lower Limit for the Energy Derivative of the Scattering
  Phase Shift}}, \href{https://doi.org/10.1103/PhysRev.98.145}{\emph{Phys.
  Rev.} {\bfseries 98} (1955) 145}.

\bibitem{Phillips:1996ae}
D.R.~Phillips and T.D.~Cohen, \emph{{How short is too short? Constraining
  contact interactions in nucleon-nucleon scattering}},
  \href{https://doi.org/10.1016/S0370-2693(96)01411-6}{\emph{Phys. Lett. B}
  {\bfseries 390} (1997) 7}
  [\href{https://arxiv.org/abs/nucl-th/9607048}{{\ttfamily nucl-th/9607048}}].

\bibitem{Hammer:2009zh}
H.W.~Hammer and D.~Lee, \emph{{Causality and universality in low-energy quantum
  scattering}},
  \href{https://doi.org/10.1016/j.physletb.2009.10.033}{\emph{Phys. Lett. B}
  {\bfseries 681} (2009) 500}
  [\href{https://arxiv.org/abs/0907.1763}{{\ttfamily 0907.1763}}].

\bibitem{Hammer:2010fw}
H.-W.~Hammer and D.~Lee, \emph{{Causality and the effective range expansion}},
  \href{https://doi.org/10.1016/j.aop.2010.06.006}{\emph{Annals Phys.}
  {\bfseries 325} (2010) 2212}
  [\href{https://arxiv.org/abs/1002.4603}{{\ttfamily 1002.4603}}].

\bibitem{Beane:2018oxh}
S.R.~Beane, D.B.~Kaplan, N.~Klco and M.J.~Savage, \emph{{Entanglement
  Suppression and Emergent Symmetries of Strong Interactions}},
  \href{https://doi.org/10.1103/PhysRevLett.122.102001}{\emph{Phys. Rev. Lett.}
  {\bfseries 122} (2019) 102001}
  [\href{https://arxiv.org/abs/1812.03138}{{\ttfamily 1812.03138}}].

\bibitem{Low:2021ufv}
I.~Low and T.~Mehen, \emph{{Symmetry from entanglement suppression}},
  \href{https://doi.org/10.1103/PhysRevD.104.074014}{\emph{Phys. Rev. D}
  {\bfseries 104} (2021) 074014}
  [\href{https://arxiv.org/abs/2104.10835}{{\ttfamily 2104.10835}}].

\bibitem{Wigner:1936dx}
E.~Wigner, \emph{{On the Consequences of the Symmetry of the Nuclear
  Hamiltonian on the Spectroscopy of Nuclei}},
  \href{https://doi.org/10.1103/PhysRev.51.106}{\emph{Phys. Rev.} {\bfseries
  51} (1937) 106}.

\bibitem{Wigner:1937zz}
E.~Wigner, \emph{{On the Structure of Nuclei Beyond Oxygen}},
  \href{https://doi.org/10.1103/PhysRev.51.947}{\emph{Phys. Rev.} {\bfseries
  51} (1937) 947}.

\bibitem{Wigner:1939zz}
E.P.~Wigner, \emph{{On Coupling Conditions in Light Nuclei and the Lifetimes of
  beta-Radioactivities}},
  \href{https://doi.org/10.1103/PhysRev.56.519}{\emph{Phys. Rev.} {\bfseries
  56} (1939) 519}.

\bibitem{Cohen:1998jr}
T.D.~Cohen and J.M.~Hansen, \emph{{Low-energy theorems for nucleon-nucleon
  scattering}}, \href{https://doi.org/10.1103/PhysRevC.59.13}{\emph{Phys. Rev.
  C} {\bfseries 59} (1999) 13}
  [\href{https://arxiv.org/abs/nucl-th/9808038}{{\ttfamily nucl-th/9808038}}].

\bibitem{deSwart:1995ui}
J.J.~de~Swart, C.P.F.~Terheggen and V.G.J.~Stoks, \emph{{The Low-energy n p
  scattering parameters and the deuteron}},  in \emph{{3rd International
  Symposium on Dubna Deuteron 95}}, 9, 1995
  [\href{https://arxiv.org/abs/nucl-th/9509032}{{\ttfamily nucl-th/9509032}}].

\bibitem{Beane:2021C}
S.R.~Beane and R.C.~Farrell, \emph{{Symmetries of the nucleon-nucleon
  $S$-matrix and effective field theory expansions}},
  \href{https://arxiv.org/abs/2112.05800}{{\ttfamily 2112.05800}}.

\bibitem{garay2019classical}
I.~Garay and S.~Robles-Pérez, \emph{Classical geodesics from the canonical
  quantisation of spacetime coordinates},
  \href{https://arxiv.org/abs/1901.05171}{{\ttfamily 1901.05171}}.

\bibitem{blau2020}
M.~Blau, \emph{Lecture notes on general relativity},  2020.

\bibitem{Beck:2019abp}
S.~Beck, B.~Bazak and N.~Barnea, \emph{{Removing the Wigner bound in
  non-perturbative effective field theory}},
  \href{https://doi.org/10.1016/j.physletb.2020.135485}{\emph{Phys. Lett. B}
  {\bfseries 806} (2020) 135485}
  [\href{https://arxiv.org/abs/1907.11886}{{\ttfamily 1907.11886}}].

\bibitem{Habashi:2020qgw}
J.B.~Habashi, S.~Sen, S.~Fleming and U.~van Kolck, \emph{{Effective Field
  Theory for Two-Body Systems with Shallow S-Wave Resonances}},
  \href{https://doi.org/10.1016/j.aop.2020.168283}{\emph{Annals Phys.}
  {\bfseries 422} (2020) 168283}
  [\href{https://arxiv.org/abs/2007.07360}{{\ttfamily 2007.07360}}].

\bibitem{PhysRevA.64.012706}
D.S.~Petrov and G.V.~Shlyapnikov, \emph{Interatomic collisions in a tightly
  confined bose gas},
  \href{https://doi.org/10.1103/PhysRevA.64.012706}{\emph{Phys. Rev. A}
  {\bfseries 64} (2001) 012706}.

\bibitem{PhysRevA.76.063610}
J.P.~Kestner and L.-M.~Duan, \emph{Effective low-dimensional hamiltonian for
  strongly interacting atoms in a transverse trap},
  \href{https://doi.org/10.1103/PhysRevA.76.063610}{\emph{Phys. Rev. A}
  {\bfseries 76} (2007) 063610}.

\bibitem{PhysRevA.85.061604}
S.K.~Baur, B.~Fr\"ohlich, M.~Feld, E.~Vogt, D.~Pertot, M.~Koschorreck et~al.,
  \emph{Radio-frequency spectra of feshbach molecules in quasi-two-dimensional
  geometries}, \href{https://doi.org/10.1103/PhysRevA.85.061604}{\emph{Phys.
  Rev. A} {\bfseries 85} (2012) 061604}.

\bibitem{PhysRevA.98.051603}
P.~Zin, M.~Pylak, T.~Wasak, M.~Gajda and Z.~Idziaszek, \emph{Quantum bose-bose
  droplets at a dimensional crossover},
  \href{https://doi.org/10.1103/PhysRevA.98.051603}{\emph{Phys. Rev. A}
  {\bfseries 98} (2018) 051603}.

\bibitem{PhysRevLett.94.210401}
H.~Moritz, T.~St\"oferle, K.~G\"unter, M.~K\"ohl and T.~Esslinger,
  \emph{Confinement induced molecules in a 1d fermi gas},
  \href{https://doi.org/10.1103/PhysRevLett.94.210401}{\emph{Phys. Rev. Lett.}
  {\bfseries 94} (2005) 210401}.

\bibitem{PhysRevA.93.063631}
S.~Lammers, I.~Boettcher and C.~Wetterich, \emph{Dimensional crossover of
  nonrelativistic bosons},
  \href{https://doi.org/10.1103/PhysRevA.93.063631}{\emph{Phys. Rev. A}
  {\bfseries 93} (2016) 063631}.

\bibitem{Beane:2018huc}
S.R.~Beane and M.~Jafry, \emph{{Dimensional crossover in non-relativistic
  effective field theory}},
  \href{https://doi.org/10.1088/1361-6455/aaf5fb}{\emph{J. Phys. B} {\bfseries
  52} (2019) 035001} [\href{https://arxiv.org/abs/1810.02868}{{\ttfamily
  1810.02868}}].

\bibitem{Braaten:2004rn}
E.~Braaten and H.W.~Hammer, \emph{{Universality in few-body systems with large
  scattering length}},
  \href{https://doi.org/10.1016/j.physrep.2006.03.001}{\emph{Phys. Rept.}
  {\bfseries 428} (2006) 259}
  [\href{https://arxiv.org/abs/cond-mat/0410417}{{\ttfamily
  cond-mat/0410417}}].

\bibitem{Beane:2010ny}
S.R.~Beane, \emph{{Ground state energy of the interacting Bose gas in two
  dimensions: An Explicit construction}},
  \href{https://doi.org/10.1103/PhysRevA.82.063610}{\emph{Phys. Rev. A}
  {\bfseries 82} (2010) 063610}
  [\href{https://arxiv.org/abs/1002.3815}{{\ttfamily 1002.3815}}].

\bibitem{Kaplan:2005es}
D.B.~Kaplan, \emph{{Five lectures on effective field theory}},  10, 2005
  [\href{https://arxiv.org/abs/nucl-th/0510023}{{\ttfamily nucl-th/0510023}}].

\bibitem{Beane:2021zvo}
S.R.~Beane, R.C.~Farrell and M.~Varma, \emph{{Entanglement minimization in
  hadronic scattering with pions}},
  \href{https://doi.org/10.1142/S0217751X21502055}{\emph{Int. J. Mod. Phys. A}
  {\bfseries 36} (2021) 2150205}
  [\href{https://arxiv.org/abs/2108.00646}{{\ttfamily 2108.00646}}].

\end{thebibliography}\endgroup

\end{document}